*Letter*

# (Quasi-)Real-Time Inversion of Airborne Time-Domain Electromagnetic Data via Artificial Neural Network

**Peng Bai [1], Giulio Vignoli [1,2,\*], Andrea Viezzoli [3], Jouni Nevalainen [4] and Giuseppina Vacca [1]**

1. Department of Civil and Environmental Engineering and Architecture, University of Cagliari, 09123 Cagliari, Italy; peng.bai@unica.it (P.B.); vaccag@unica.it (G.V.)
2. GEUS—Geological Survey of Denmark and Greenland, 8000 Aarhus, Denmark
3. Aarhus Geophysics ApS, 8000 Aarhus, Denmark; andrea.viezzoli@aarhusgeo.com
4. Oulu Mining School, University of Oulu, 90014 Oulu, Finland; jouni.nevalainen@oulu.fi
\* Correspondence: gvignoli@unica.it

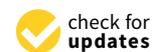

**Abstract:** The possibility to have results very quickly after, or even during, the collection of electromagnetic data would be important, not only for quality check purposes, but also for adjusting the location of the proposed flight lines during an airborne time-domain acquisition. This kind of readiness could have a large impact in terms of optimization of the Value of Information of the measurements to be acquired. In addition, the importance of having fast tools for retrieving resistivity models from airborne time-domain data is demonstrated by the fact that Conductivity-Depth Imaging methodologies are still the standard in mineral exploration. In fact, they are extremely computationally efficient, and, at the same time, they preserve a very high lateral resolution. For these reasons, they are often preferred to inversion strategies even if the latter approaches are generally more accurate in terms of proper reconstruction of the depth of the targets and of reliable retrieval of true resistivity values of the subsurface. In this research, we discuss a novel approach, based on neural network techniques, capable of retrieving resistivity models with a quality comparable with the inversion strategy, but in a fraction of the time. We demonstrate the advantages of the proposed novel approach on synthetic and field datasets.

## 1. Introduction

Airborne time-domain electromagnetic (ATEM) data have been collected for decades for mineral prospection [1–4]. Probably, the Stanmac-McPhar ATEM system—tested during the summer of 1948 in Canada—can be considered the first application of this methodology [5]. Since then, many technological advancements have been implemented [6–8] and several new acquisition systems have been built. Nowadays, numerous different helicopter and fixed-wing acquisition systems are available on the market; a few examples among the most internationally known are: MEGATEM/GEOTEM [9,10], MULTIPULSE [11], TEMPEST [12,13], SkyTEM [14], Spectrum [15], VTEM [16,17], EQUATOR [18], and AeroTEM [19]. In addition, the capabilities of the available instrumentation allowed to move from the mere mineral target detection to more sophisticated geological modelling [20–22] and groundwater mapping [23,24] applications.

Not only the equipment, but also the data processing and inversion strategies have gone through continuous advancements. In this respect, just to mention an example, stacking of the recorded transient curves via moving windows with widths that are time-gate dependent [25,26] can increase the lateral resolution at shallow, while enhancing the signal-to-noise ratio at depth (where, in any case, the physics of the methodology leads to larger observation footprints). On the other hand, novel inversion strategies made it possible to enforce spatial coherence [27,28] and allowed the retrieval of (pseudo-)3D resistivity distribution even by using relatively simple 1D forward modelling [29,30].

Clearly, increasingly often, inversion schemes based on actual 3D forward modelling are used [31]. However, these approaches are still too computationally expensive to be routinely applied. Actually, every time preserving the maximum lateral resolution and obtaining fast results is a priority, imaging approaches [32–34], rather than inversion strategies, are still preferred. The results from imaging and inversion approaches can be significantly different (in this respect, see, for example, the comparison in [35]). Of course, the overall effectiveness of the approach highly depends on the final goals of the survey.

In this research, we discuss an approach based on data-driven Machine Learning (ML) algorithms and, specifically, Artificial Neural Networks (ANNs) that potentially combines the advantages of both imaging and inversion as it allows near real-time reconstructions of the resistivity distribution of the subsurface with an accuracy comparable with the physically based inversion.

ANNs are clearly not new in the processing of geophysical data [36–40]. However, the attempts to apply them to ATEM observations are limited to the data processing [41] and geological interpretation of the geophysical results [42]. In this paper, we discuss the application of ANNs for the reconstruction of the pseudo-3D electrical resistivity distribution in the subsurface from the data collected during typical ATEM surveys. We test the proposed workflow on both synthetic and field datasets and prove that the corresponding results are comparable to an inversion based on a full-nonlinear 1D forward modelling algorithm. Moreover, the main advantage of our approach is that the geophysical model of the subsurface is obtained almost instantaneously on a standard laptop.

These levels of accuracy, reconstruction speed, and flexibility could pave the road to real-time adjustments of survey planning. In seismic exploration, it is not unusual that dedicated optimizations of the survey design are performed in order to save significant resources and, at the same time, enhance the value-of-information of the collected data [43–45]. By having at hand the capability to invert the data in real-time, we can think of survey plans adapting while the data are collected. Potentially this can save the time and efforts connected to subsequent data acquisitions (in the ATEM case, maybe, ground based) as follow-ups of the original survey.

## 2. Methods

ATEM data are usually inverted by minimizing an objective functional that consists of the summation of a data misfit and a regularization term. Hence, the objective functional to be minimized is often formalized as follows:

$$P^{(\lambda)}(\mathbf{d},\mathbf{m}) = \| \mathbf{W}_d \, (\mathbf{d}_{obs} - F(\mathbf{m})) \|_{L2}^2 + \lambda \, s(\mathbf{m}), \tag{1}$$

in which (i) $\mathbf{d}_{obs}$ is the vector of the observations; (ii) $\mathbf{m}$ is the vector of the model parameters; (iii) $F$ is the forward modelling operator mapping the model $\mathbf{m}$ into the data space; $F$ takes into account the physics of the process and the characteristics of the acquisition system [46]; (iv) $\mathbf{W}_d$ is the data weighting matrix taking into account the uncertainty in the measurements; (v) $s(\mathbf{m})$ is the regularization term incorporating the prior knowledge about the resistivity model to be reconstructed; (vi) the multiplier $\lambda$ controls the balance between the importance given to the data with respect to the prior information.

In the deterministic scheme we are investigating here, in order to have a term of comparison to effectively assess the performances of the alternative approach based on ANN, we consider a one-dimensional model parameterization. Hence, $\mathbf{m}$ (and consistently also $F$) is based on the assumption that (locally) the subsurface is not varying laterally. Therefore, even when data and model sections are discussed, each individual data sounding, and each associated model, is handled independently from the adjacent ones. More specifically, whereas the forward modelling $F$ is always one dimensional, there is a connection between the neighboring models imposed through the regularization term. In this respect, concerning the stabilizer choice, we adopt the probably most common option of $s(\mathbf{m})$ being equal to the minimum gradient stabilizer.

$$s(\mathbf{m}) = \parallel \nabla \mathbf{m} \parallel_{L2}^2. \tag{2}$$

Hence, despite the conductivity distribution is considered locally 1D, the stabilizer acts both along the vertical (*z*) and the horizontal (*x*) direction, promoting solutions that are laterally coherent (without being truly 2D/3D). This is the essence of the so-called spatially constrained inversion [28,47].

Moreover, in the 1D deterministic inversion scheme we are using, the value of $\lambda$ is calculated in order to guarantee a chi-squared value

$$\chi^2 = \left(\frac{1}{N_d}\right) \parallel \mathbf{W}_d \left(\mathbf{d}_{obs} - F(\mathbf{m})\right) \parallel_{L2}^2 \tag{3}$$

approximately equal 1 (with $N_d$ being the number of time gates) [48,49].

The ANN is built in order to perform a similar task with respect to the minimization of objective functional in Equation (1).

ANNs use continuous and differentiable activation functions at each unit of the network, which makes the network output (**m**) a continuous and differentiable function of the network input (**d**); this, in turn, leads to the possibility of defining a continuous and differentiable error function for the evaluation of the difference between the network output and the target output. Consequently, the error function can be minimized over a training set using a relatively simple gradient-based procedure. Hence, the problem of building an effective ANN to map the recorded measurements into resistivity vector is reduced to the minimization of an error functional:

$$E(\mathbf{w}) = \parallel K(\mathbf{D}, \mathbf{w}) - \mathbf{M} \parallel_{L2}^2, \tag{4}$$

in which **D** and **M** consist of the elements of the (data, model) couples ($\mathbf{d}_t$, $\mathbf{m}_t$) constituting the Training Dataset (TD) [50]. Of course, in this case, the minimization aims at finding the optimal weights **w** of the connections between the network units. Thus, the ANN *K* is found via the minimization with respect to **w**. Once *K* is built based on the TD, it can be applied to the elements $\mathbf{d}_{obs}$ of the observed dataset to infer the corresponding conductivity models **m**. In this respect, it is worth noting that the retrieved *K*—and, therefore, the corresponding final resistivity distribution obtained via the application of *K* on the observed data—depends on the selection of the TD. ML approaches are based on the stationarity assumption: the couples in the TD and in the solution dataset need to be independent and identically distributed (i.i.d.) random variables. In this sense, TD formalizes the available prior information about the studied system. Consistently, the TD should be selected in order to be representative of the targets (therefore, coherent with our expectations about the geology to be reconstructed) [50,51]. Data stationarity and TD's representativeness are very well-known issues of ML [52]. In a further attempt to reconcile the ANN approach and the (regularized) deterministic inversion, we could think about the selection of the conductivity models for the development of the TD as some sort of regularization: the solution provided by the ANN cannot be too different from the models (and the associated data) used to train the ANN. Hence, for example, the TD should be based on the prior (geological) knowledge available about the investigated area. This might sound tautological, but it is actually the key point of regularization theory (and, clearly, also of ML approaches).

In the present paper, the ANN consists of a multilayer perceptron with (i) an input layer with 54 (i.e., the number of time gates) units; (ii) three hidden layers with, respectively, 100, 500, 200 units; and (iii) an output layer characterized by 30 (i.e., the number of conductivity model parameters) units.

As TD we took the $\mathbf{d}_t$ data generated via the forward modelling *F* for each of the 1D resistivity models $\mathbf{m}_t$ making up a realistic resistivity section (Figure 1). It is important to stress that, despite the apparent lateral coherence of the 1D model, the elements of the TD are, indeed, handled as independent soundings and resistivity models. Plotting the TD data and the models as 2D sections made it easier to assess the representativeness of the (geologically informed) training dataset with respect to the actual measurements to be inverted.

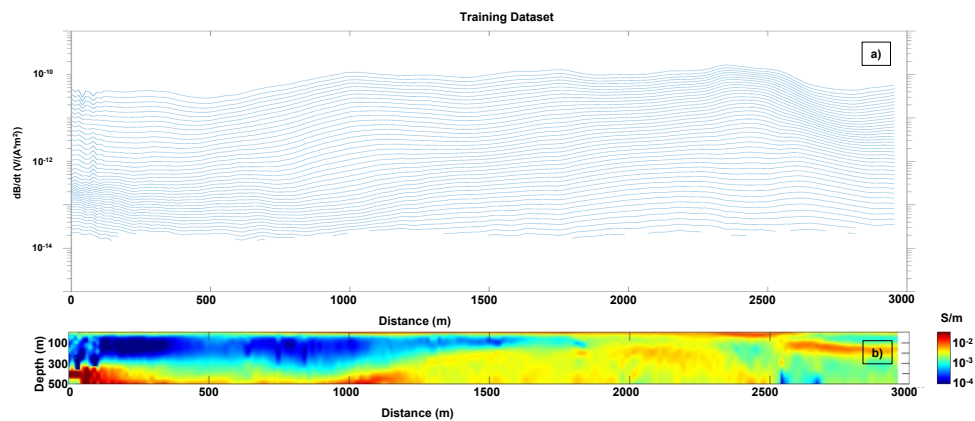

**Figure 1.** Portion of the Training Dataset. Panel (**a**) shows the airborne time-domain electromagnetic (ATEM) responses associated with the 1D models constituting the conductivity section in panel (**b**).

It is also important to highlight that the TD used in the present research is based on the technical specifications of the particular system used for the experimental data collection. Thus, the $\mathbf{d}_t$'s are calculated from the corresponding $\mathbf{m}_t$'s by using, for example, the waveform and time gates provided by the contractor together with the survey measurements.

In total, the utilized TD consisted of around 12,000 ($\mathbf{d}_t$, $\mathbf{m}_t$) couples (a sample of which is plotted in Figure 1). In the training phase, a multi-start approach has been adopted to minimize the effect of local minima of the error functional. Additionally, following a standard procedure [53], the optimal number of epochs was selected by studying the error functional value when applied on validation subsets [54].

Differently from the 1D deterministic inversion case in Equation (1) (in which the stabilizing terms connects adjacent 1D models), in the inversion performed through $K$, no lateral information is included, and the individual soundings are inverted separately. The inclusion of this further piece of knowledge would be surely beneficial (if available) and should be included in future developments.

## 3. Results

### 3.1. Synthetic Test

In order to assess the effectiveness of the ANN approach, we applied the neural network (based on the previously discussed TD) to a known verification dataset.

Figure 2 shows the true conductivity sections whose 1D models were used to generate the noise-free synthetic data to be inverted. Therefore, in short, and by using a neural network lingo, Figure 2 (together with its associated data) is our verification dataset.

Figure 3 consists of the conductivity sections reconstructed via the proposed ANN. In turn, the inferred conductivities (Figure 3) have been used to calculate their associated electromagnetic response; the comparison between the original synthetic data and the calculated response is shown, model-by-model, with a red dot (red axis on the right in Figure 3). From this data misfit estimation, it is clear that the conductivity distribution recovered by the neural network is generally compatible with the inverted data within 4%.

Considering the retrieved conductivity distribution, the ANN reconstruction captures almost all the features present in the original model. In addition, Figure 3 demonstrates that the proposed approach is quite robust as it retrieves the lateral coherence of the conductivity sections despite the individual models are inverted separately. A quantitative assessment of the model agreement between the reconstructed and the original model can be done through the Figure 4 showing, in the log-scale, the ratio between the ANN reconstruction and the true model. In general, the values in Figure 4 are around one, demonstrating the overall accuracy of the ANN reconstruction. The areas in Figure 4 characterized by major discrepancies between the ANN solution and the true model are generally localized at depth (where, in any case, because of the physics of the method, the sensitivity of the data to the conductivity values is lower) and on the right side of the conductivity sections. This is not surprising if we look at the electromagnetic responses. Regarding this, Figure 5 shows the original data (blue lines) compared to the calculated measurements (red lines) for each of the sections in Figures 2 and 3; it is clear that many of the soundings on the right side of the sections are characterized by a smaller number of time gates (indeed, to simulate more realistic conditions, in several of the original soundings, the late time gates have been removed, mimicking what often happens with field noisy observations). Of course, with a reduced number of time gates, the depth at which the conductivity affects the data values is shallower. This is consistent with the larger model misfit on the right side of the panels in Figure 4.

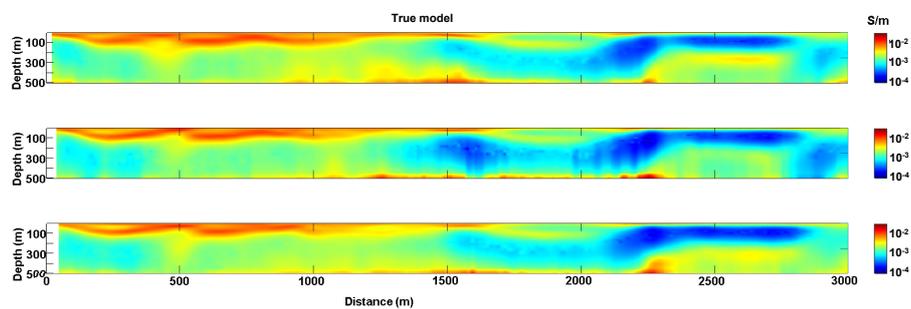

**Figure 2.** The verification dataset. The individual 1D conductivity models of these sections have been used to generate the noise-free synthetic data to be subsequently inverted with the Artificial Neural Networks (ANN) discussed in the section "Methods".

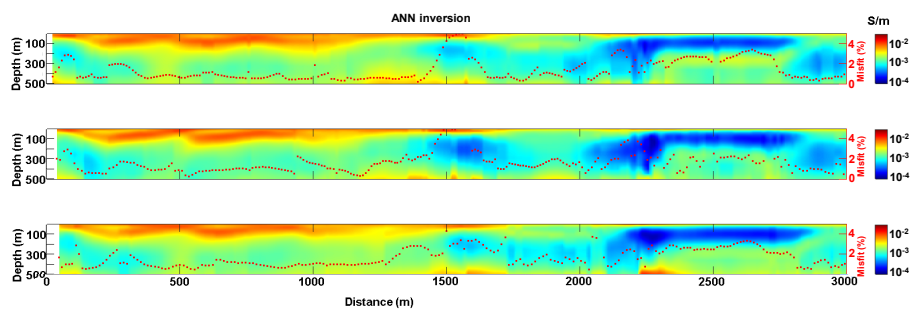

**Figure 3.** The inversion results obtained by applying the ANN trained on the training dataset (TD) in Figure 1 to the data generated by the conductive models in Figure 2. The data misfit between the calculated and the original measurements is shown for each individual 1D model location as a red dot (the corresponding axis is on the right in red).

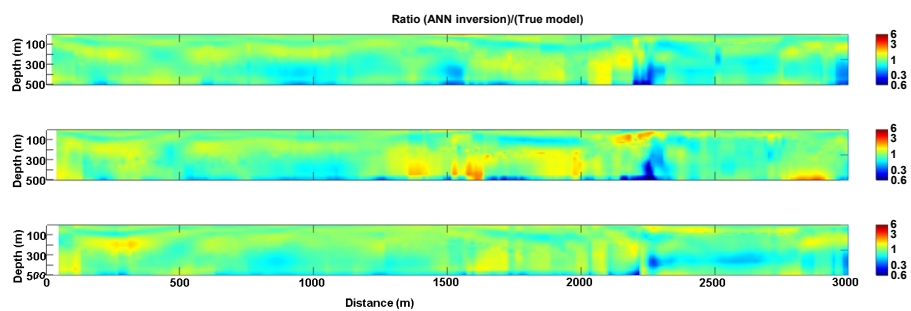

**Figure 4.** The ratio between the conductivity models in Figure 3 (the ANN result) and in Figure 2 (the true conductivity distribution).

*3.2. Field Example*

The Sakatti deposit is located 15 km north of Sodankylä, Finland; it is rich in Cu-Ni-PGE (Platinum Group Elements) minerals [55] and has been selected as one of the test sites of the EU-funded project INFACT aiming at the development of cutting-edge technologies for mineral exploration [56]. Within the framework of the INFACT project, time-domain electromagnetic data have been acquired by Geotech using a VTEM system [16,17].

In this section, we compare the results obtained with the ANN—already used for the previous synthetic test and trained on the TD in Figure 1—against a more traditional 1D deterministic inversion based on the forward modelling utilized for simulating, for example, the responses in Figure 5.

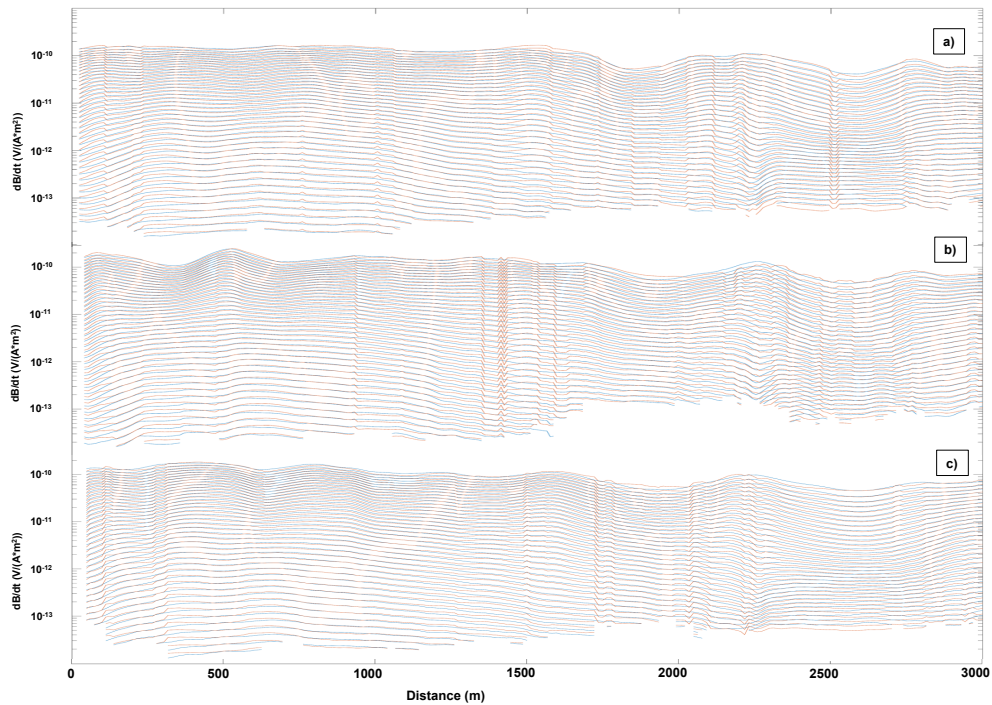

**Figure 5.** Comparison between the synthetic data from the true model in Figure 2 (blue lines) and the calculated data from the conductivity sections retrieved by the ANN in Figure 3 (red lines). The data in the panels (**a**–**c**) in the present figure correspond to the conductivity sections shown in the three panels in each of the Figures 2 and 3.

Therefore, Figure 6 demonstrates that the inversion performed via the developed ANN can infer reasonable 1D models whose responses fit the observation within a 5% threshold (for each model, the data misfit value is represented as a red dot, and the associated red axis is on the right side of the panel).

When the ANN result is compared with the corresponding 1D deterministic inversion in Figure 7, it is possible to see that the "traditional" deterministic inversion with vertical and lateral smooth constrained is often superior in fitting the data (the data misfit is generally below 2%, as it is clearly visible from the red dots representing the data misfit). Figure 8 might be helpful in quantitatively evaluating the differences between the two results as it shows the ratio between the different solutions; it worth noting how the larger discrepancies between the two solutions occur where the data fitting of the deterministic inversion is larger (e.g., between 1850 and 2400 m) and/or in areas characterized by high resistivity values. Therefore, the areas in which also the deterministic inversion has difficulties in fitting the observations and that are characterized by relatively high resistivity values are those where the differences with the ANN solution are more pronounced. This is in agreement with the fact that, in general, ATEM methods have difficulties in accurately distinguish between different high resistivity values.

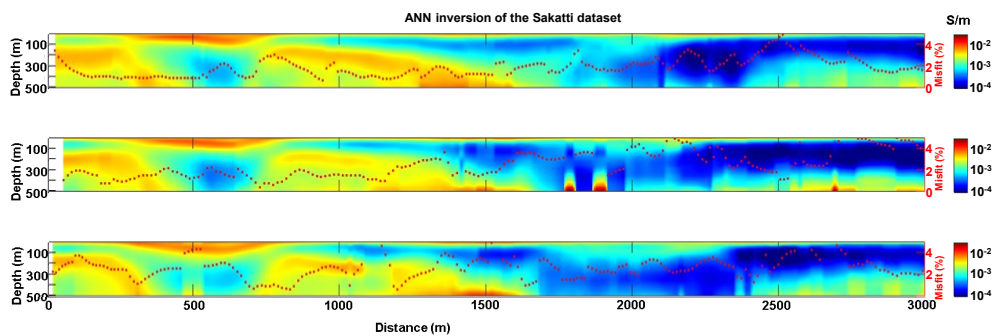

**Figure 6.** The ANN inversion of the field data.

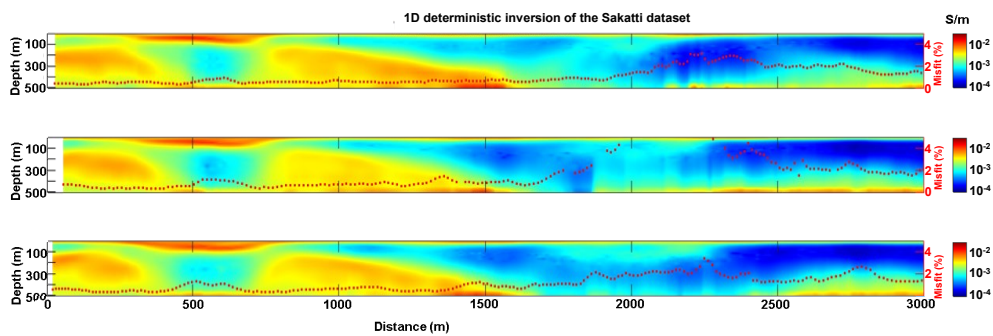

**Figure 7.** The 1D deterministic inversion of the field data.

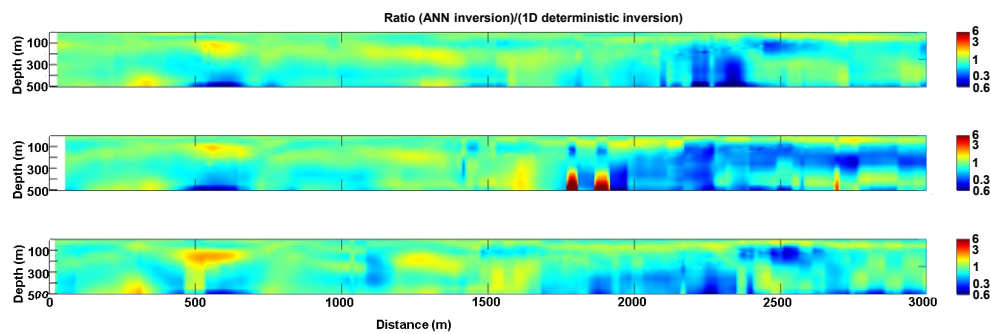

**Figure 8.** The ratio between the conductivity models in Figure 6 (the ANN inversion result) and in Figure 7 (the deterministic inversion result).

## 4. Discussion

Clearly, the indubitable advantages of the deterministic inversion come at a price: the ANN inversion takes approximately 24 s to invert the entire Sakatti dataset (consisting of 14,346 soundings with 54 time gates each) by using a standard laptop (equipped with an Intel Core i5-8250U processor), whereas hours (so an amount of time of the order of magnitude of $10^4$ s) are necessary to perform the same task by using the 1D deterministic approach and a 64-CPU server.

To be fair, it is true that the training phase—crucial for the development of the ANN—requires several hours. Despite that, we believe that the proposed workflow has at least a few main pros:

1. It can allow the optimization of the survey design while the acquisition of the ATEM is on-going. In fact, the development of an effective training dataset and the associate ANN can be performed before the survey—or it can be even based on the outcomes from the first flight(s) of the survey if the area is assumed to be relatively "stationary"—and, once the ANN is available, reliable results can be almost instantaneously obtained just after each flight. In turn, this can lead to real-time rearrangements of the original tentative survey plans in order to maximize the VoI (Value of Information) of the measurements to be further collected [57].

2. The ANN speed can be extremely useful for effective Quality Check (QC) of the data during the survey.
3. The availability of a good starting model (derived from the ANN inversion) can be used to speed-up the 1D deterministic inversion by reducing the number of iterations.

Of course, if, for producing the final results, post-processing analyses are necessary (e.g., in 3D environments), the same will be true also when adopting the proposed ANN approach: ANN based on a 1D forward modelling approach cannot guarantee better results compared with the corresponding deterministic inversion; it can only provide solutions of similar quality within a fraction of the time and by using cheaper computational tools.

Moreover, clearly, at least in principle, the presented ANN scheme can be extended in order to also include, for example, induce polarization (IP) effects: it is a matter of incorporating them in the forward modelling algorithm used in the development of the TD. However, as before, we cannot expect the ANN to solve all the issues connected IP in ATEM data. Reasonably, we can only expect to solve the same problem, but much faster.

## 5. Conclusions

We present a novel approach to the inversion of airborne time-domain electromagnetic data based on neural networks.

We demonstrate the effectiveness of the proposed inversion strategy by testing it on both synthetic and field data. Based on these outcomes, we conclude that the proposed neural network approach is capable of retrieving the conductivity distribution of the subsurface from the measurements collected by the airborne geophysical system with an accuracy that is largely comparable with the most commonly used (in the academia and in the industry) inversion strategies and that relies on 1D deterministic inversion approaches. These results are particularly noticeable as the neural network inversion takes only a fraction of the time required for the deterministic inversion (a few seconds versus hours).

The performances of the neural network discussed in the paper can be potentially enhanced in terms of data fitting via data augmentation techniques expanding the TD and building more accurate models, provided that effective ways to generate artificial data getting closer to the behavior of the test dataset can be found [58,59].

In addition, an aspect that has not been investigated here is the dependence on the training dataset; in future works, studying the robustness of the result as a function of the training dataset would be extremely relevant: after all, the definition of the proper training dataset is a way to include prior (geological) information into the inversion.

Clearly, the neural network strategy discussed in the paper deals with each sounding independently and does not make use of the possible available knowledge concerning the lateral coherence of the targets; it is a pity not to exploit these additional pieces of information in order to get even more effective results. In this perspective, pseudo-2D/3D approaches should be explored as well.

The dramatic speed up of the inversion by means of the application of the neural network (seconds vs. hours, on a standard laptop) potentially paves the road to on-the-fly inversion with possible applications on real-time survey design optimizations. On the other hand, in the most conservative scenario, the discussed neural network inversion can serve as a starting model for faster deterministic inversions and/or as a QC tool during the data collection phases.

**Author Contributions:** Conceptualization, P.B. and G.V. (Giulio Vignoli); methodology, P.B. and G.V. (Giulio Vignoli); software, P.B. and G.V. (Giulio Vignoli); validation, P.B., G.V. (Giulio Vignoli), A.V. and J.N.; formal analysis, P.B. and G.V. (Giulio Vignoli); investigation, P.B., G.V. (Giulio Vignoli) and A.V.; resources, G.V. (Giulio Vignoli); data curation, P.B., G.V. (Giulio Vignoli), and A.V.; writing—original draft preparation, P.B. and G.V. (Giulio Vignoli); writing—review and editing, P.B., G.V. (Giulio Vignoli), A.V., J.N. and G.V. (Giuseppina Vacca); visualization, P.B. and G.V. (Giulio Vignoli); supervision, G.V. (Giulio Vignoli); project administration, G.V. (Giulio Vignoli) and G.V. (Giuseppina Vacca); funding acquisition, G.V. (Giulio Vignoli) and G.V. (Giuseppina Vacca). All authors have read and agreed to the published version of the manuscript.

**Funding:** This work was partially supported: by the initiative PON-RI 2014-2020, Asse I "Capitale Umano"—Azione I.1 "Dottorati innovativi con caratterizzazione industriale Ciclo XXXIII"—project: "GEOPROBARE: stochastic inversion of time-domain electromagnetic data"; by the INFACT project (https://www.infactproject.eu/) funded by the European Union's Horizon 2020 research and innovation programme (grant agreement no. 776487); by the initiative POR-FESR Sardegna 2014-2020, Asse I Azione I.1.3 "Creare opportunità di lavoro favorendo la competitività delle imprese"—project: "Tecnologie di CARatterizzazione Monitoraggio e Analisi per il ripristino e la bonifica (CARMA)"; by the project "GETHERE" (RAS/FBS grant-number: F71/17000190002).

**Acknowledgments:** Thanks are due to Anglo American for making the data available for this study. Moreover, P.B. and G.V. (Giulio Vignoli) are grateful to Antonello Sanna, Roberto Ricciu and Giuseppe Desogus of the University of Cagliari (DICAAR) for providing most of the computer resources necessary to perform these tests. All authors are deeply thankful to Battista Biggio of the University of Cagliari (PRA Lab) for his invaluable comments and suggestions; without Battista, this research would not have been successful.

**Conflicts of Interest:** The authors declare no conflict of interest.

## References

1. Zhdanov, M.S.; Alfouzan, F.A.; Cox, L.; Alotaibi, A.; Alyousif, M.; Sunwall, D.; Endo, M. Large-scale 3D modeling and inversion of multiphysics airborne geophysical data: A case study from the Arabian Shield, Saudi Arabia. *Minerals* **2018**, *8*, 271. [CrossRef]
2. Witherly, K. The quest for the Holy Grail in mining geophysics: A review of the development and application of airborne EM systems over the last 50 years. *Lead. Edge* **2000**, *19*, 270–274. [CrossRef]
3. Alfouzan, F.A.; Alotaibi, A.M.; Cox, L.H.; Zhdanov, M.S. Spectral Induced Polarization Survey with Distributed Array System for Mineral Exploration: Case Study in Saudi Arabia. *Minerals* **2020**, *10*, 769. [CrossRef]
4. Cudahy, T. Mineral mapping for exploration: An Australian journey of evolving spectral sensing technologies and industry collaboration. *Geosciences* **2016**, *6*, 52. [CrossRef]
5. Fountain, D. Airborne electromagnetic systems-50 years of development. *Explor. Geophys.* **1998**, *29*, 1–11. [CrossRef]
6. Fraser, D.C. Resistivity mapping with an airborne multicoil electromagnetic system. *Geophysics* **1978**, *43*, 144–172. [CrossRef]
7. Smith, R.S.; Fountain, D.; Allard, M. The MEGATEM fixed-wing transient EM system applied to mineral exploration: A discovery case history. *First Break* **2003**, *21*, 73–77.
8. Rasmussen, S.; Nyboe, N.S.; Mai, S.; Larsen, J.J. Extraction and use of noise models from transient electromagnetic data. *Geophysics* **2018**, *83*, E37–E46. [CrossRef]
9. Smith, R.S.; Lee, T.J. The moments of the impulse response: A new paradigm for the interpretation of transient electromagnetic data. *Geophysics* **2002**, *67*, 1095–1103. [CrossRef]
10. Annan, A.P.; Lookwood, R. An application of airborne GEOTEM in Australian conditions. *Explor. Geophys.* **1991**, *22*, 5–12. [CrossRef]
11. Chen, T.; Hodges, G.; Miles, P. MULTIPULSE–high resolution and high power in one TDEM system. *Explor. Geophys.* **2015**, *46*, 49–57. [CrossRef]
12. Macnae, J. Improving the accuracy of shallow depth determinations in AEM sounding. *Explor. Geophys.* **2004**, *35*, 203–207. [CrossRef]
13. Peters, G.; Street, G.; Kahimise, I.; Hutchins, D. Regional TEMPEST survey in north-east Namibia. *Explor. Geophys.* **2015**, *46*, 27–35. [CrossRef]
14. Sørensen, K.I.; Auken, E. SkyTEM-A new high-resolution helicopter transient electromagnetic system. *Explor. Geophys.* **2004**, *35*, 191–199. [CrossRef]
15. Leggatt, P.B.; Klinkert, P.S.; Hage, T.B. The Spectrem airborne electromagnetic system—Further developments. *Geophysics* **2000**, *65*, 1976–1982. [CrossRef]
16. Legault, J.M.; Izarra, C.; Prikhodko, A.; Zhao, S.; Saadawi, E.M. Helicopter EM (ZTEM–VTEM) survey results over the Nuqrah copper–lead–zinc–gold SEDEX massive sulphide deposit in the Western Arabian Shield, Kingdom of Saudi Arabia. *Explor. Geophys.* **2015**, *46*, 36–48. [CrossRef]
17. Kwan, K.; Prikhodko, A.; Legault, J.M.; Plastow, G.C.; Kapetas, J.; Druecker, M. VTEM airborne EM, aeromagnetic and gamma-ray spectrometric data over the Cerro Quema high sulphidation epithermal gold deposits, Panama. *Explor. Geophys.* **2016**, *47*, 179–190. [CrossRef]

18. Karshakov, E.V.; Podmogov, Y.G.; Kertsman, V.M.; Moilanen, J. Combined Frequency Domain and Time Domain Airborne Data for Environmental and Engineering Challenges. *J. Environ. Eng. Geophys.* **2017**, *22*, 1–11. [CrossRef]
19. Boyko, W.; Paterson, N.R.; Kwan, K. AeroTEM characteristics and field results. *Lead. Edge* **2001**, *20*, 1130–1138. [CrossRef]
20. Jørgensen, F.; Møller, R.R.; Nebel, L.; Jensen, N.P.; Christiansen, A.V.; Sandersen, P.B. A method for cognitive 3D geological voxel modelling of AEM data. *Bull. Eng. Geol. Environ.* **2013**, *72*, 421–432. [CrossRef]
21. Siemon, B.; Ibs-von Seht, M.; Frank, S. Airborne electromagnetic and radiometric peat thickness mapping of a bog in Northwest Germany (Ahlen-Falkenberger Moor). *Remote Sens.* **2020**, *12*, 203. [CrossRef]
22. Høyer, A.S.; Jørgensen, F.; Sandersen, P.B.E.; Viezzoli, A.; Møller, I. 3D geological modelling of a complex buried-valley network delineated from borehole and AEM data. *J. Appl. Geophys.* **2015**, *122*, 94–102. [CrossRef]
23. Siemon, B.; Christiansen, A.V.; Auken, E. A review of helicopter-borne electromagnetic methods for groundwater exploration. *Near Surf. Geophys.* **2009**, *7*, 629–646. [CrossRef]
24. Sapia, V.; Viezzoli, A.; Jørgensen, F.; Oldenborger, G.A.; Marchetti, M. The impact on geological and hydrogeological mapping results of moving from ground to airborne TEM. *J. Environ. Eng. Geophys.* **2014**, *19*, 53–66. [CrossRef]
25. Reninger, P.A.; Martelet, G.; Perrin, J.; Dumont, M. Processing methodology for regional AEM surveys and local implications. *Explor. Geophys.* **2020**, *51*, 143–154. [CrossRef]
26. Liu, R.; Guo, R.; Liu, J.; Liu, Z. An Efficient Footprint-Guided Compact Finite Element Algorithm for 3-D Airborne Electromagnetic Modeling. *IEEE Geosci. Remote Sens. Lett.* **2019**, *16*, 1809–1813. [CrossRef]
27. Auken, E.; Christiansen, A.V. Layered and laterally constrained 2D inversion of resistivity data. *Geophysics* **2004**, *69*, 752–761. [CrossRef]
28. Vignoli, G.; Fiandaca, G.; Christiansen, A.V.; Kirkegaard, C.; Auken, E. Sharp spatially constrained inversion with applications to transient electromagnetic data. *Geophys. Prospect.* **2014**, *63*, 243–255. [CrossRef]
29. Ley-Cooper, A.Y.; Viezzoli, A.; Guillemoteau, J.; Vignoli, G.; Macnae, J.; Cox, L.; Munday, T. Airborne electromagnetic modelling options and their consequences in target definition. *Explor. Geophys.* **2015**, *46*, 74–84. [CrossRef]
30. Viezzoli, A.; Auken, E.; Munday, T. Spatially constrained inversion for quasi 3D modelling of airborne electromagnetic data–an application for environmental assessment in the Lower Murray Region of South Australia. *Explor. Geophys.* **2009**, *40*, 173–183. [CrossRef]
31. Cox, L.H.; Wilson, G.A.; Zhdanov, M.S. 3D inversion of airborne electromagnetic data. *Geophysics* **2012**, *77*, WB59–WB69. [CrossRef]
32. Wolfgram, P.; Karlik, G. Conductivity-depth transform of GEOTEM data. *Explor. Geophys.* **1995**, *26*, 179–185. [CrossRef]
33. Macnae, J.; King, A.; Stolz, N.; Osmakoff, A.; Blaha, A. Fast AEM data processing and inversion. *Explor. Geophys.* **1998**, *29*, 163–169. [CrossRef]
34. Huang, H.; Rudd, J. Conductivity-depth imaging of helicopter-borne TEM data based on a pseudolayer half-space model. *Geophysics* **2008**, *73*, F115–F120. [CrossRef]
35. Dzikunoo, E.A.; Vignoli, G.; Jørgensen, F.; Yidana, S.M.; Banoeng-Yakubo, B. New regional stratigraphic insights from a 3D geological model of the Nasia sub-basin, Ghana, developed for hydrogeological purposes and based on reprocessed B-field data originally collected for mineral exploration. *Solid Earth* **2020**, *11*, 349–361. [CrossRef]
36. Brykov, M.N.; Petryshynets, I.; Pruncu, C.I.; Efremenko, V.G.; Pimenov, D.Y.; Giasin, K.; Sylenko, S.A.; Wojciechowski, S. Machine learning modelling and feature engineering in seismology experiment. *Sensors* **2020**, *20*, 4228. [CrossRef] [PubMed]
37. Núñez-Nieto, X.; Solla, M.; Gómez-Pérez, P.; Lorenzo, H. GPR signal characterization for automated landmine and UXO detection based on machine learning techniques. *Remote Sens.* **2014**, *6*, 9729–9748. [CrossRef]
38. Rymarczyk, T.; Kłosowski, G.; Kozłowski, E. A non-destructive system based on electrical tomography and machine learning to analyze the moisture of buildings. *Sensors* **2018**, *18*, 2285. [CrossRef]
39. Yuan, S.; Liu, J.; Wang, S.; Wang, T.; Shi, P. Seismic waveform classification and first-break picking using convolution neural networks. *IEEE Geosci. Remote Sens. Lett.* **2018**, *15*, 272–276. [CrossRef]

40. Van der Baan, M.; Jutten, C. Neural networks in geophysical applications. *Geophysics* **2000**, *65*, 1032–1047. [CrossRef]
41. Andersen, K.K.; Kirkegaard, C.; Foged, N.; Christiansen, A.V.; Auken, E. Artificial neural networks for removal of couplings in airborne transient electromagnetic data. *Geophys. Prospect.* **2016**, *64*, 741–752. [CrossRef]
42. Gunnink, J.L.; Bosch, J.H.A.; Siemon, B.; Roth, B.; Auken, E. Combining ground-based and airborne EM through Artificial Neural Networks for modelling glacial till under saline groundwater conditions. *Hydrol. Earth Syst. Sci.* **2012**, *16*, 3061. [CrossRef]
43. Bhuiyan, M.; Sacchi, M. Optimization for sparse acquisition. In SEG Technical Program Expanded Abstracts. In Proceedings of the Society of Exploration Geophysicists 85th Annual Meetings and International Expositions, New Orleans, LA, USA, 18–23 October 2015.
44. Latiff, A.H.A.; Ghosh, D.P.; Latiff, N.M.A.A. Optimizing acquisition geometry in shallow gas cloud using particle swarm optimization approach. *Int. J. Comput. Intell. Syst.* **2017**, *10*, 1198–1210. [CrossRef]
45. Curtis, A. Optimal design of focused experiments and surveys. *Geophys. J. Int.* **1999**, *139*, 205–215. [CrossRef]
46. Auken, E.; Christiansen, A.V.; Kirkegaard, C.; Fiandaca, G.; Schamper, C.; Behroozmand, A.A.; Binley, A.; Nielsen, E.; Effersø, F.; Christensen, N.B.; et al. An overview of a highly versatile forward and stable inverse algorithm for airborne, ground-based and borehole electromagnetic and electric data. *Explor. Geophys.* **2015**, *46*, 223–235. [CrossRef]
47. Vignoli, G.; Sapia, V.; Menghini, A.; Viezzoli, A. Examples of improved inversion of different airborne electromagnetic datasets via sharp regularization. *J. Environ. Eng. Geophys.* **2017**, *22*, 51–61. [CrossRef]
48. Vignoli, G.; Deiana, R.; Cassiani, G. Focused inversion of vertical radar profile (VRP) traveltime data. *Geophysics* **2012**, *77*, H9–H18. [CrossRef]
49. Vignoli, G.; Guillemoteau, J.; Barreto, J.; Rossi, M. Reconstruction, with tunable sparsity levels, of shear-wave velocity profiles from surface wave data. *Geophys. J. Int.*. under review.
50. Bishop, C.M. *Pattern Recognition and Machine Learning*; Springer: New York, NY, USA, 2006.
51. Alpaydin, E. *Introduction to Machine Learning*; MIT Press: Cambridge, MA, USA, 2004.
52. Han, D.; Lee, J.; Im, J.; Sim, S.; Lee, S.; Han, H. A novel framework of detecting convective initiation combining automated sampling, machine learning, and repeated model tuning from geostationary satellite data. *Remote Sens.* **2019**, *11*, 1454. [CrossRef]
53. Liu, Y.; Starzyk, J.A.; Zhu, Z. Optimized Approximation Algorithm in Neural Networks without Overfitting. *IEEE Trans. Neural Netw.* **2008**, *19*, 983–995. [PubMed]
54. Duda, R.O.; Hart, P.E.; Stork, D.G. *Pattern Classification*; John Wiley & Sons: New York, NY, USA, 2012.
55. Brownscombe, W.; Ihlenfeld, C.; Coppard, J.; Hartshorne, C.; Klatt, S.; Siikaluoma, J.K.; Herrington, R.J. The Sakatti Cu-Ni-PGE sulfide deposit in northern Finland. In *Mineral Deposits of Finland*; Lahtinen, R., O'Brien, H., Maier, W.D., Eds.; Elsevier: Amsterdam, The Netherlands, 2015; pp. 211–252.
56. Kesselring, M.; Wagner, F.; Kirsch, M.; Ajjabou, L.; Gloaguen, R. Sustainable Test Sites for Mineral Exploration: Development of Sustainable Test Sites and Knowledge Spillover for Industry. *Sustainability* **2020**, *12*, 2016. [CrossRef]
57. Eidsvik, J.; Bhattacharjya, D.; Mukerji, T. Value of information of seismic amplitude and CSEM resistivity. *Geophysics* **2008**, *73*, R59–R69. [CrossRef]
58. Zhang, F.; Chan, P.P.; Biggio, B.; Yeung, D.S.; Roli, F. Adversarial feature selection against evasion attacks. *IEEE Trans. Cybern.* **2015**, *46*, 766–777. [CrossRef] [PubMed]
59. Biggio, B.; Russu, P.; Didaci, L.; Roli, F. Adversarial biometric recognition: A review on biometric system security from the adversarial machine-learning perspective. *IEEE Signal Process. Mag.* **2015**, *32*, 31–41. [CrossRef]